\documentclass[prd,aps,floats,floatfix,nofootinbib,preprint]{revtex4-1}
\usepackage{amsmath}
\usepackage{amsfonts}
\usepackage{graphicx}
\usepackage{slashed}
\usepackage{dcolumn}
\usepackage{bm}
\usepackage{amssymb}
\usepackage{latexsym}
\usepackage{color}
\usepackage{longtable}

\newcommand{\ordEW}{\mathcal{O}(\alpha_{\scriptscriptstyle EM}^6)}
\newcommand{\ordQCD}{\mathcal{O}(\alpha_{\scriptscriptstyle EM}^4
  \alpha_{\scriptscriptstyle S}^2)}

\def\blap#1{\vbox to\z@{#1\vss}}

\newcommand{\ifb}{\mbox{ fb}^{-1}}
\newcommand{\tbn}[1]{Tab.~\ref{#1}}
\newcommand{\eqn}[1]{Eq.(\ref{#1})}
\newcommand{\fig}[1]{Fig.~\ref{#1}}

\newcommand{\ra}{\rightarrow}
\newcommand{\GeV}{\mbox{ ${\mathrm GeV}$}}

\begin{document}

\title{Probing Near-Conformal Technicolor through Weak Boson Scattering}

\author{D. Buarque Franzosi}
\email[]{diogo.buarque@uclouvain.be}
\author{R. Foadi}
\email[]{roshan.foadi@uclouvain.be}
\affiliation{Centre for Cosmology, Particle Physics and Phenomenology (CP3)
Chemin du Cyclotron 2, Universit\'e catholique de Louvain, Belgium}

\begin{abstract}
The recently observed boson at 125 GeV could be a light composite scalar from near-conformal technicolor dynamics: a technicolor Higgs. If this is the case, unitarization of longitudinal weak boson scattering amplitudes, which is due to exchanges of the Higgs and spin-one vector technimesons, is expected to occur in a strong regime, with saturation of the unitarity bounds. This implies that $pp\to VVjj$ processes, where $V$ is either a $W$ or a $Z$ boson, are enhanced, relative to the standard model. We show that this allows probing near-conformal technicolor for couplings and masses of the spin-one resonances which are not directly accessible for direct Drell-Yan production.
\end{abstract}

\maketitle

\section{Introduction}\label{sec:intro}
Recently the ATLAS and CMS collaborations have announced the discovery of a new boson with the approximate mass of 125 GeV, and suggested that this might be the long sought Higgs boson~\cite{ATLAS:2012ae,Chatrchyan:2012tx}. If this is confirmed by future data, it will be extremely important to accurately measure the Higgs coupling to the particles of the standard model (SM). The $g_{hWW}$ and $g_{hZZ}$ couplings are of particular importance, because of the role they play in the unitarization of longitudinal weak boson scattering amplitudes. Recall that for a given $2\to2$ scattering amplitude ${\cal M}$ the $J$-th partial wave projection is defined by
\begin{equation}
a_J\equiv \frac{1}{32\pi}\int_{-1}^1 dx\ P_J(x)\ {\cal M},\quad J=0,1,2,\dots\ ,
\end{equation}
where $x$ is the cosine of the scattering angle, and $P_J(x)$ is the $J$-th Legendre polynomial. For any value of $J$, unitarity demands $a_J$ to lie on a unit circle centered on $i/2$, the Argand circle. In particular this implies the bounds
\begin{equation}
-\frac{1}{2}\leq{\rm Re}\ a_J\leq\frac{1}{2} \ .
\label{eq:unitbound}
\end{equation}
Consider for instance the $W_L^+ W_L^-\to W_L^+ W_L^-$ scattering. At large values of the center-of-mass (CM) energy the amplitude grows like
\begin{equation}
{\cal M}(W_L^+ W_L^-\to W_L^+ W_L^-) \sim \left(1-r_h^2\right)\frac{-2(1+x)E^2}{v^2} \ ,
\label{eq:WWh}
\end{equation}
where $v$ is the electroweak vev, $v\simeq 246$ GeV, and
\begin{equation}
r_h\equiv\frac{g_{hWW}}{g_{hWW}^{\rm SM}} \ .
\end{equation}
If $r_h$ is well below one, for instance, the $a_0(W_L^+ W_L^-\to W_L^+ W_L^-)$ projection undergoes unitarity loss at a few TeVs~\cite{Veltman:1976rt,Veltman:1977fy,Lee:1977yc,Lee:1977eg,Passarino:1985ax,Passarino:1990hk}. If this is the case, new particles, not heavier than a few TeVs, must come into play and unitarize the amplitude.

It is often assumed that a light Higgs boson is associated with perturbative dynamics. This implies that, at high energy, the partial wave projections of the longitudinal weak boson scattering amplitudes have a real part approaching a small and constant value, far from the unitarity bounds of Eq.~(\ref{eq:unitbound}). The reason for this is that the imaginary part of an amplitude is due to loop corrections, which, in a perturbative regime, are small. Thus in order for the $a_J$ projections to lie on the Argand circle, also the real part has to be small.  Therefore, the terms growing like $E^2$ must vanish. If $r_h\neq 1$, this implies that new states must supply the right-hand side of Eq.~(\ref{eq:WWh}),
\begin{equation}
{\cal M}(W_L^+ W_L^-\to W_L^+ W_L^-) \sim \left(1-r_h^2-\sum_\rho r_\rho^2\right)\frac{-2(1+x)E^2}{v^2} \ ,
\label{eq:WWhrho}
\end{equation}
in such a way that the coefficient of the term growing like $E^2$ vanishes:
\begin{equation}
r_h^2+\sum_\rho r_\rho^2 = 1 \ .
\label{eq:sum}
\end{equation}
This, for example, is the scenario conjectured in~\cite{Falkowski:2012vh} and~\cite{Bellazzini:2012tv}.

Despite the common lore, we believe it possible for a light Higgs to coexist with strong dynamics at the unitarization scale. This occurs if the Higgs boson is a composite scalar of new strong dynamics, technicolor (TC), which is responsible for breaking the electroweak symmetry. Naively the natural mass scale for the lightest TC scalar singlet is of the order of 1 TeV, as obtained by scaling up the lightest scalar isosinglet of quantum chromo-dynamics (QCD), the $\sigma$ meson. However, in viable TC theories naive QCD scaling-up does not hold, as there are additional mechanisms which lead to a large suppression of the mass of the lightest scalar, relative to the naive estimate. First, viable TC theories are expected to be near-conformal in a range of energies above the chiral symmetry breaking scale. To see this, recall that the SM fermions acquire mass by interacting with the TC vacuum through a spontaneously broken gauge interaction, extended technicolor (ETC)~\cite{Dimopoulos:1979es,Eichten:1979ah}. This, however, introduces unwanted flavor-changing neutral currents, which can only be eliminated by raising the mass of the ETC gauge bosons. Doing so would result in unacceptably small SM fermion masses, unless the TC vacuum is enhanced, from the TC to the ETC scale, by near-conformal dynamics and a large anomalous dimension of the TC vacuum~\cite{Holdom:1981rm}. If TC dynamics feel the presence of a nearby fixed point, the mass of the lightest scalar is expected to be reduced, relative to a QCD-like spectrum. The reason for this is that TC theories with a spontaneously broken scale invariance are expected to feature a massless scalar singlet in the spectrum, the technidilaton~\cite{Yamawaki:1985zg,Bando:1986bg}. In asymptotically-free TC theories scale invariance is only approximate, and the technidilatonic Higgs boson is expected to acquire a small dynamical mass~\cite{Yamawaki:1985zg,Bando:1986bg,Dietrich:2005jn,Appelquist:2010gy}. By small we mean smaller than the natural mass scale of the light technihadrons, which is of the order of 1 TeV.

Second, the Higgs mass in TC, as well as in the SM, is modified by radiative corrections due to gauge interactions and ETC-induced Yukawa interactions. As in the SM, in TC the radiative corrections to the Higgs mass are quadratic in the cutoff. However, unlike the SM, a theory of technihadrons has a physical cutoff, the scale of compositeness. The latter is estimated to be of the order of $4\pi F_\Pi$~\cite{Manohar:1983md}, where $F_\Pi$ is the ``technipion'' decay constant, analogous to the pion decay constant of QCD. If the fermionic sector of TC consists of $N_{\rm TD}$ weak ``technidoublets'', each contributing an equal amount to $v^2$, then $F_\Pi=v/\sqrt{N_{\rm TD}}$. Since the cutoff is physical, the cutoff-dependence of the radiative corrections to the Higgs mass is also physical, and, being quadratic, large. The dominant contribution to $\delta M_h^2$ is negative: it arises from the top loop, and is of the order of $\left(600\ {\rm GeV}\right)^2$~\cite{FoadiFrandsenSannino}.

Finally, the TC Higgs can become lighter by "geometric scaling", as the number of technicolors and/or weak technidoublets is increased. The TC states become lighter as $N_{\rm TD}$ increases because, as mentioned in the last paragraph, the square of the electroweak vev receives contribution from $N_{\rm TD}$ weak pairs. If $v$ is to be kept fixed at 246 GeV, this forces the whole TC spectrum to scale like $1/\sqrt{N_{\rm TD}}$. The scaling as a function of the number of technicolors, $N_{\rm TC}$, is much more subtle. The large-$N_{\rm TC}$ limit of $SU(N_{\rm TC})$ was first obtained by 't-Hooft with technifermions in the fundamental representation~\cite{'tHooft:1973jz}. This allows rescaling from QCD, which corresponds to $N_{\rm TC}=3$. Corrigan and Ramond noted that scaling-up QCD is also possible with technifermions in the two-index antisymmetric representation, as the latter coincides with the anti-fundamental representation for $N_{\rm TC}=3$~\cite{Corrigan:1979xf}. The 't Hooft and Corrigan-Ramond large-$N_{\rm TC}$ limits lead to qualitatively very different spectra. In particular, only $\overline{Q}Q$ scalars, in the 't Hoof limit, with $v$ fixed at 246 GeV, become lighter at large values of $N_{\rm TC}$, whereas multiquark states do not exist at $N_{\rm TC}\to\infty$. On the other hand, in the Corrigan-Ramond limit scalar masses are always expected to scale down like $\sqrt{2/N_{\rm TC}(N_{\rm TC}-1)}$. Large $N_{\rm TC}$ scaling can be obtained for other higher-dimensional representations, even though a comparison with QCD is no longer possible. For instance, in the two-index symmetric representation scalars masses are expected to scale like $\sqrt{2/N_{\rm TC}(N_{\rm TC}+1)}$, in the large $N_{\rm TC}$ limit~\cite{Sannino:2009za}.

For strongly coupled dynamics at the unitarization scale, we neither expect sum rules like the one of Eq.~(\ref{eq:sum}) to hold, nor the partial wave projections of longitudinal weak boson scattering amplitudes to have a small real part. Instead, we expect the latter to quickly grow (with either an overall plus or minus sign) and oscillate between the unitarity bounds -1/2 and 1/2. This is indeed the scenario of pion-pion scattering in QCD~\cite{Harada:1995dc}. Because of this, in models with strong unitarization we expect enhancement of $p p \to V V j j$ processes at the LHC, where each $V$ denotes either a $W$ or a $Z$ boson~\cite{Bagger:1993zf,Bagger:1995mk}. Here we analyze this in the context of near-conformal TC, with a light composite Higgs in the spectrum, and vector meson dominance (VMD). The interesting aspect of this analysis is that it allows to probe TC for masses and coupling of the spin-one resonances lying outside the LHC reach for significant production through Drell-Yan (DY) processes. In fact in models of VMD the coupling of the light SM fermions to the spin-one technimesons is suppressed by $g/\tilde{g}$, where $g$ is the weak coupling and $\tilde{g}$ is the spin-one technimeson coupling. Thus increasing $\tilde{g}$ reduces sensitivity to DY production~\cite{Belyaev:2008yj,Andersen:2011nk,Frandsen:2011hj}. On the other hand, increasing $\tilde{g}$ does not lead to a reduction of the cross section for $p p \to V V j j$, as the partial wave projections of longitudinal weak boson scattering amplitudes are always expected to oscillate between -1/2 and 1/2. Furthermore, DY production of spin-one technimesons becomes increasingly kinematically forbidden as the latter become heavier. On the other hand, in $p p \to V V j j$ processes the strength of the amplitude does not decrease as the technimeson masses become larger. In fact the technimesons are expected to unitarize the weak boson scattering amplitudes, no matter what their mass is\footnote{Of course unitarity imposes bounds on the mass of the TC spin-one resonances, and is therefore inconsistent to take the latter to be unreasonably heavy.}. Our goal is therefore to show that the discovery of TC through $p p \to V V j j$ processes is complementary to DY production of spin-one technimesons.

The remainder of this note is organized as follows. In Sec.~\ref{sec:unit} we review aspects of unitarity of longitudinal weak boson scattering in near-conformal TC, with a light Higgs and a technirho isospin triplet unitarizing the amplitudes. In Sec.\ref{sec:vector} we analyze the $p p \to V V j j$ processes at the LHC, and the potential for discovery of near-conformal TC. Finally, in Sec.~\ref{sec:conclusions} we offer our conclusions.
\section{Unitarity in near-conformal technicolor}\label{sec:unit}
The model we use as a template for our analysis is near-conformal TC with an $SU(2)_L\times SU(2)_R\to SU(2)_V$ chiral symmetry breaking pattern. By gauging the electroweak subgroup of $SU(2)_L\times SU(2)_R$ this becomes a model of dynamical electroweak symmetry breaking. The near conformal behavior of the underlying gauge theory is reflected, in the effective theory, by the presence of a light Higgs, which we choose to take as the chiral partner of the technipions eaten by the $W$ and $Z$ bosons, and by a near degenerate set of spin-one vector and axial-vector isotriplets. Such a model has been termed ``vanilla TC'', and its phenomenology analyzed in~\cite{Belyaev:2008yj,Andersen:2011nk,Frandsen:2011hj}. Unitarity of longitudinal weak boson scattering amplitudes in TC has instead been studied in~\cite{Foadi:2008xj}.
\subsection{Lagrangian and parametrization}
We denote the normalized $SU(2)$ generators with $T^a$, $a=1,2,3$, where $2T^a$ are the Pauli matrices. Employing a  linear representation for the chiral group leads to the effective Lagrangian\footnote{In~\cite{Belyaev:2008yj,Andersen:2011nk,Frandsen:2011hj} $\alpha_1$, $\alpha_2$, and $\beta$ are termed $r_2$, $r_3$, and $s$, respectively.}
\begin{eqnarray}
{\cal L} &=& {\cal L}_{\overline{\rm SM}}
-\frac{1}{2}\ {\rm Tr} \left[L_{\mu\nu}L^{\mu\nu}+R_{\mu\nu}R^{\mu\nu}\right]
+\frac{1}{2}\ {\rm Tr}\left[D_\mu M D^\mu M^\dagger\right]
+m^2\ {\rm Tr} \left[C_{L\mu}^2+C_{R\mu}^2\right] \nonumber \\
&-&\tilde{g}^2\ \alpha_1\ {\rm Tr}\left[C_{L\mu} M C_R^\mu M^\dagger\right]
-i\ \frac{\tilde{g}\ \alpha_2}{4}\ {\rm Tr}\left[C_{L\mu}\left(M D^\mu M^\dagger-D^\mu M M^\dagger\right)
+ C_{R\mu}\left(M^\dagger D^\mu M-D^\mu M^\dagger M\right)\right] \nonumber \\
&+&\frac{\tilde{g}^2\ \beta}{4}\ {\rm Tr} \left[C_{L\mu}^2+C_{R\mu}^2\right] {\rm Tr}\left[M M^\dagger\right]
+\frac{\mu^2}{2}\ {\rm Tr}\left[M M^\dagger\right]
-\frac{\lambda}{4}\ {\rm Tr}\left[M M^\dagger\right]^2
+ {\cal L}_{\overline{\psi}\psi-{\rm resonance}} + {\cdots} \ ,
\label{eq:L}
\end{eqnarray}
where the ellipses denote higher dimensional terms, suppressed by inverse powers of the cutoff $\Lambda\simeq 4\pi v$. Here ${\cal L}_{\overline{\rm SM}}$ is the SM Lagrangian without the Higgs sector and the Yukawa interactions, whereas ${\cal L}_{\overline{\psi}\psi-{\rm resonance}}$ contains the ETC-induced interactions of the SM fermions with the TC vacuum (mass terms) and the TC resonances. The field-strength tensors $L_{\mu\nu}$ and $R_{\mu\nu}$ are associated to the spin-one resonance triplets $L_\mu\equiv L_\mu^a T^a$ and $R_\mu\equiv R_\mu^a T^a$, respectively, with coupling $\tilde{g}$. The fields $C_{L\mu}$ and $C_{R\mu}$ are defined by
\begin{equation}
C_{L\mu}\equiv L_\mu-\frac{g}{\tilde{g}}W_\mu \ , \quad
C_{R\mu}\equiv R_\mu-\frac{g^\prime}{\tilde{g}}B_\mu \ ,
\end{equation}
where $W_\mu\equiv W_\mu^a T^a$ and $B_\mu$ are the electroweak boson gauge eigenstates. These definitions follow from the fact that the fields $L_\mu$ and $R_\mu$ are taken to transform like $W_{\mu}$ and $B_\mu$, respectively, under the electroweak gauge group. In the limit of zero electroweak gauge couplings, and because of chiral symmetry breaking, the technimeson mass eigenstates are not $L_\mu$ and $R_\mu$, but the vector and axial linear combinations,
\begin{equation}
V_\mu\equiv \frac{L_\mu+R_\mu}{\sqrt2} \ , \quad
A_\mu\equiv \frac{L_\mu-R_\mu}{\sqrt2} \ .
\end{equation}
When the electroweak gauge couplings are switched on, $V_\mu$ and $A_\mu$ further mix with $W_\mu$ and $B_\mu$ to form mass eigenstates. The spin-zero matrix $M$ transforms like the bifoundamental of $SU(2)_L\times SU(2)_R$, and is defined by
\begin{equation}
M\equiv \frac{1}{\sqrt2}\left[\hat{v}+h + 2 i \pi^a T^a\right] \ ,
\end{equation}
where $\pi^a$ are the technipions ``eaten'' by the SM weak bosons, and $\hat{v}$ is the chiral symmetry breaking vev. The latter is not quite equal to $v\equiv 246$ GeV. In fact the mixing term $A_\mu^a\partial\pi^a$, which arises from the term proportional to $\alpha_2$ in the effective Lagrangian, must be removed prior to quantization. This requires a shift of the $A_\mu^a$ fields and a tree-level renormalization of the technipions. Thus $\hat{v}$ approaches $v\equiv 246$ GeV only for large values of the axial technimeson mass, and/or small values of the coupling $\tilde{g}$. The precise relation between $v$ and $\hat{v}$ is
\begin{equation}
v^2 = \hat{v}^2\left[1-\frac{\alpha_2^2\ \tilde{g}^2\ \hat{v}^2}{8\ M_A^2}\right] \ .
\label{eq:FP}
\end{equation}
Upon higgsing, the spin-zero field $M$ contributes to the mass of the spin-one vector and axial triplets. For zero electroweak gauge couplings this gives
\begin{equation}
M_V^2 = m^2 + \frac{(\beta-\alpha_1)\ \tilde{g}^2\ \hat{v}^2}{4} \ , \quad
M_A^2 = m^2 + \frac{(\beta+\alpha_1)\ \tilde{g}^2\ \hat{v}^2}{4} \ .
\end{equation}

In addition to the SM parameters, the Lagrangian of Eq.~(\ref{eq:L}) contains five additional parameters: $m$, $\tilde{g}$, $\alpha_1$, $\alpha_2$, and $\beta$. In a QCD-like TC theory we could impose both Weinberg sum rules (WSRs)~\cite{Weinberg:1967kj} on the vector and axial resonances to eliminate two of these parameters. However in near-conformal TC we only expect the first WSR to approximately hold~\cite{Appelquist:1998xf}, which leaves us with four input parameters. In~\cite{Belyaev:2008yj,Andersen:2011nk,Frandsen:2011hj} these were taken to be the mass of the axial technimeson, the coupling $\tilde{g}$, the Lagrangian parameter $\beta$, and the contribution to the Peskin-Takeuchi $S$ parameter~\cite{Peskin:1991sw} from the axial and vector technimesons, ${\cal S}$. The latter is given by~\cite{Foadi:2007ue}
\begin{equation}
{\cal S} = \frac{8\pi}{\tilde{g}^2}\left[1-\left(1-\frac{\alpha_2\ \tilde{g}^2\ \hat{v}^2}{4\ M_A^2}\right)^2\right] \ .
\label{eq:SVA}
\end{equation}
\subsection{Unitarity of longitudinal weak boson scattering amplitudes}\label{sec:unitscatt}
At high energies the longitudinal weak boson scattering amplitudes approach the scattering amplitudes of the corresponding eaten Goldstone bosons~\cite{Lee:1977yc,Chanowitz:1985hj}. We adopt the principle of local unitarization: at a given CM energy $\sqrt{s}$, the scattering amplitudes are unitarized by leading-order interactions, and exchanges of states with mass not heavier than $\sqrt{s}$. In our TC model the isospin invariant amplitude is unitarized by Higgs and spin-one vector exchanges,
\begin{eqnarray}
A(s,t,u) = \left[\frac{1}{v^2}-\frac{3 g_{V\pi\pi}^2}{M_V^2}\right]s
-\frac{g_{h\pi\pi}^2}{v^2}\frac{s^2}{s-M_h^2}
-g_{V\pi\pi}^2\left[\frac{s-u}{t-M_V^2}+\frac{s-t}{u-M_V^2}\right] \ .
\label{eq:invariant}
\end{eqnarray}
Here $g_{V\pi\pi}$ is the coupling of the vector technimeson to the Goldstone bosons, and $g_{h\pi\pi}$ is the relevant linear combination of the Higgs couplings to the eaten Goldstone bosons in units of $1/v$~\cite{Foadi:2008xj}. The parameters of Eq.~(\ref{eq:invariant}), $M_V$, $g_{V\pi\pi}$, and $g_{h\pi\pi}$, can be computed in terms of the input parameters $M_A$, $\tilde{g}$, ${\cal S}$, and $\beta$~\cite{Foadi:2008xj}. This gives
\begin{equation}
g_{V\pi\pi} = \frac{\tilde{g}\ M_V^2\ {\cal S}}{8\pi\sqrt2\ v^2}\ , \quad
g_{h\pi\pi} = \frac{v}{\hat{v}}\left[1-\frac{m^2 (1-\beta)}{M_A^2}\left(\frac{\hat{v}^2}{v^2}-1\right)\right] \ ,
\label{eq:gVpph}
\end{equation}
and
\begin{equation}
M_V = M_A\sqrt{1-\frac{\tilde{g}^2{\cal S}}{8\pi}+\frac{\tilde{g}^2 v^2}{2 M_A^2}} \ .
\end{equation}

The isospin projections of the invariant amplitude are~\cite{Bagger:1993zf}
\begin{eqnarray}
I_0 &=& 3 A(s,t,u)+A(t,s,u)+A(u,t,s)\ , \nonumber \\
I_1 &=& A(t,s,u)-A(u,t,s)\ , \nonumber \\
I_2 &=& A(t,s,u)+A(u,t,s) \ .
\end{eqnarray}
The Goldstone boson scattering amplitudes can in turn be expressed as functions of the isospin projections
\begin{eqnarray}
&&{\cal M}(W_L^+ W_L^-\to Z_L Z_L) = \frac{1}{3}\left[I_0-I_2\right]\ , \nonumber \\
&&{\cal M}(W_L^+ W_L^-\to W_L^+ W_L^-) = \frac{1}{6}\left[2I_0+3I_1+I_2\right]\ , \nonumber \\
&&{\cal M}(W_L^\pm Z_L\to W_L^\pm Z_L) = \frac{1}{2}\left[I_0+I_2\right]\ , \nonumber \\
&&{\cal M}(W_L^\pm W_L^\pm \to W_L^\pm W_L^\pm) = I_2\ , \nonumber \\
&&{\cal M}(Z_L Z_L\to Z_L Z_L) = \frac{1}{3}\left[I_0+2I_2\right] \ .
\end{eqnarray}
For the $W_L^+ W_L^-\to W_L^+ W_L^-$ scattering, and taking into account corrections due to nonzero electroweak couplings, this gives
\begin{equation}
{\cal M}(W_L^+ W_L^-\to W_L^+ W_L^-) \sim \left[1-g_{h\pi\pi}^2-\frac{3 g_{V\pi\pi}^2 v^2}{M_V^2}
+{\cal O}\left(\frac{g^2}{\tilde{g}^2}\right)\right]\frac{-2(1+x)E^2}{v^2} \ ,
\label{eq:WWhV}
\end{equation}
whence, comparing with Eq~(\ref{eq:WWhrho}),
\begin{eqnarray}
r_h^2 = g_{h\pi\pi}^2 + {\cal O}\left(\frac{g^2}{\tilde{g}^2}\right) \ , \quad
r_V^2 = \frac{3 g_{V\pi\pi}^2 v^2}{M_V^2} + {\cal O}\left(\frac{g^2}{\tilde{g}^2}\right)\ .
\end{eqnarray}
\begin{figure}
\includegraphics[width=3in]{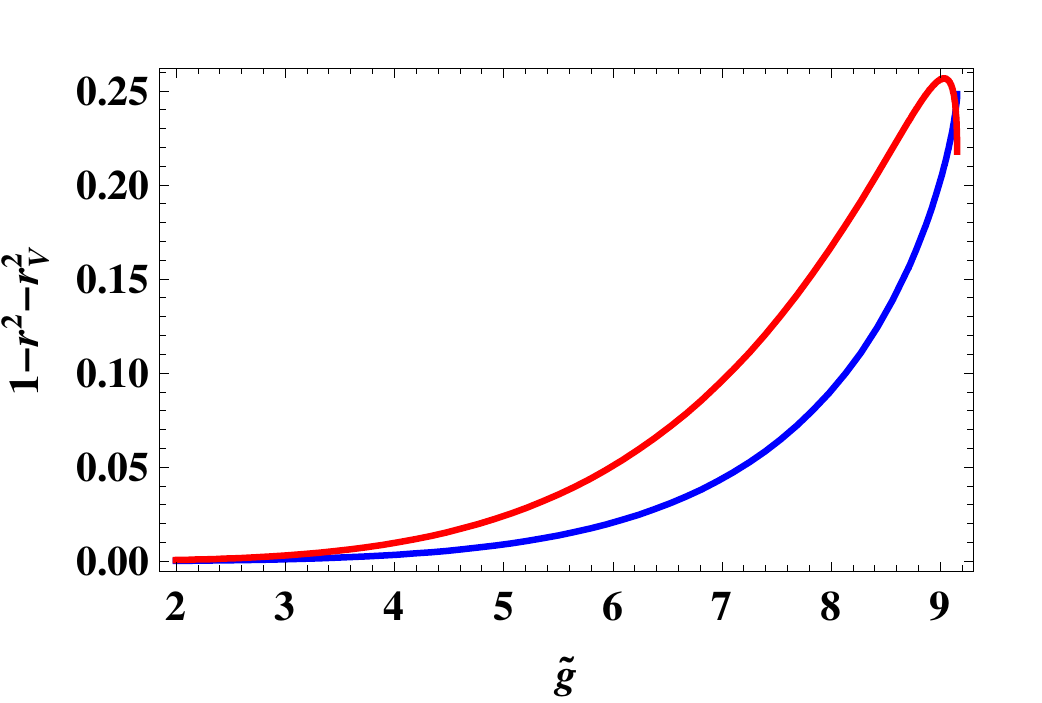}
\includegraphics[width=3in]{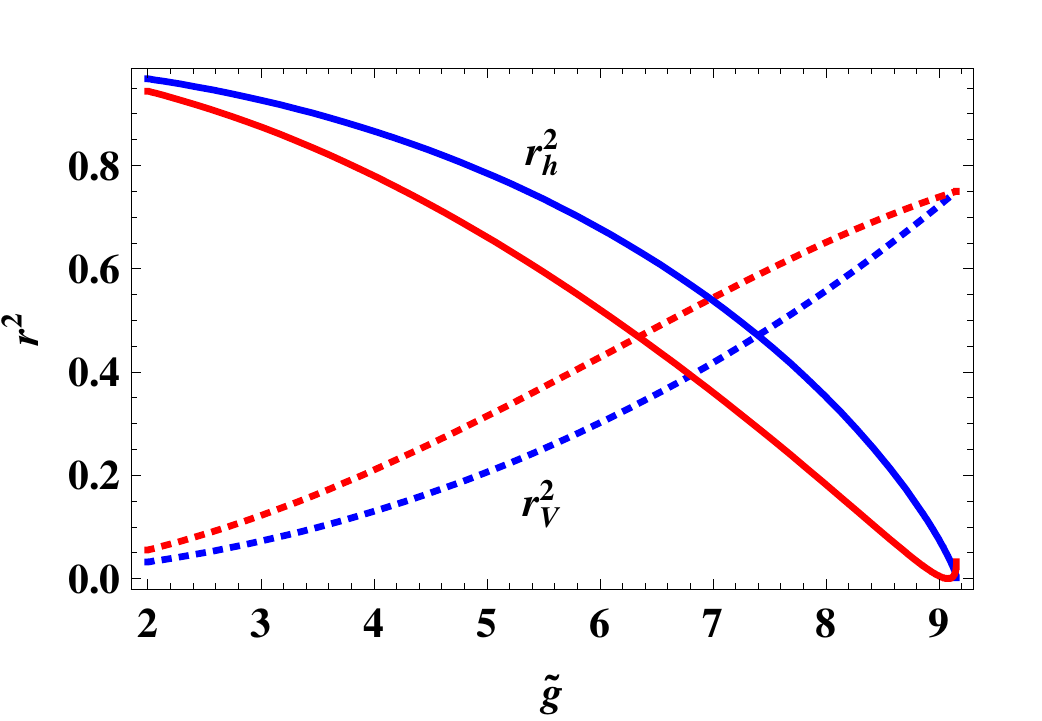}
\caption{Left: coefficient of the term growing like $E^2$ in $W_L^+ W_L^-\to W_L^+ W_L^-$ scattering (the quantity between parenthesis in Eq.~(\ref{eq:WWhV})), for ${\cal S}=0.3$, $\beta=0$, and up to small electroweak corrections. The blue lower curve is for $M_A=1.5$ TeV, whereas the red upper curve is for $M_A=2.0$ TeV. Right: the corresponding behavior of $r_h^2$ (solid curves) and $r_V^2$ (dotted curves).}
\label{fig:sumrule}
\end{figure}
In general we expect the sum rule of Eq.~(\ref{eq:sum}) to be violated, $r_h^2+r_V^2\neq 1$. The behavior of $1-r_h^2-r_V^2$ is shown in Fig.~\ref{fig:sumrule} (left), up to small electroweak corrections, for ${\cal S}=0.3$ and $\beta=0$. The blue lower curve is for $M_A=1.5$ TeV, whereas the red upper curve is for $M_A=2.0$ TeV. Eq.~(\ref{eq:SVA}) implies an upper bound on $\tilde{g}$,
\begin{equation}
\tilde{g}\leq\sqrt{\frac{8\pi}{{\cal S}}} \ ,
\end{equation}
which for ${\cal S}=0.3$ is approximately $\tilde{g}\lesssim 9.1$. In Fig.~\ref{fig:sumrule} (right) we plot the corresponding behavior of $r_h^2$ (solid curves) and $r_V^2$ (dotted curves). It might seem unreasonable that $r_h$, which is the $g_{hWW}$ coupling in SM units, decreases with increasing $M_A$, since this appears to be the decoupling limit. However this is not so, because, in addition to $v$, we are also keeping ${\cal S}$ fixed. Then inspection of Eqs.~(\ref{eq:FP}) and (\ref{eq:SVA}) shows that $\alpha_2$ and $\hat{v}$ must grow, when $M_A$ grows, and $g_{h\pi\pi}$, from Eq.~(\ref{eq:gVpph}), decreases.

\begin{figure}
\includegraphics[width=2.3in]{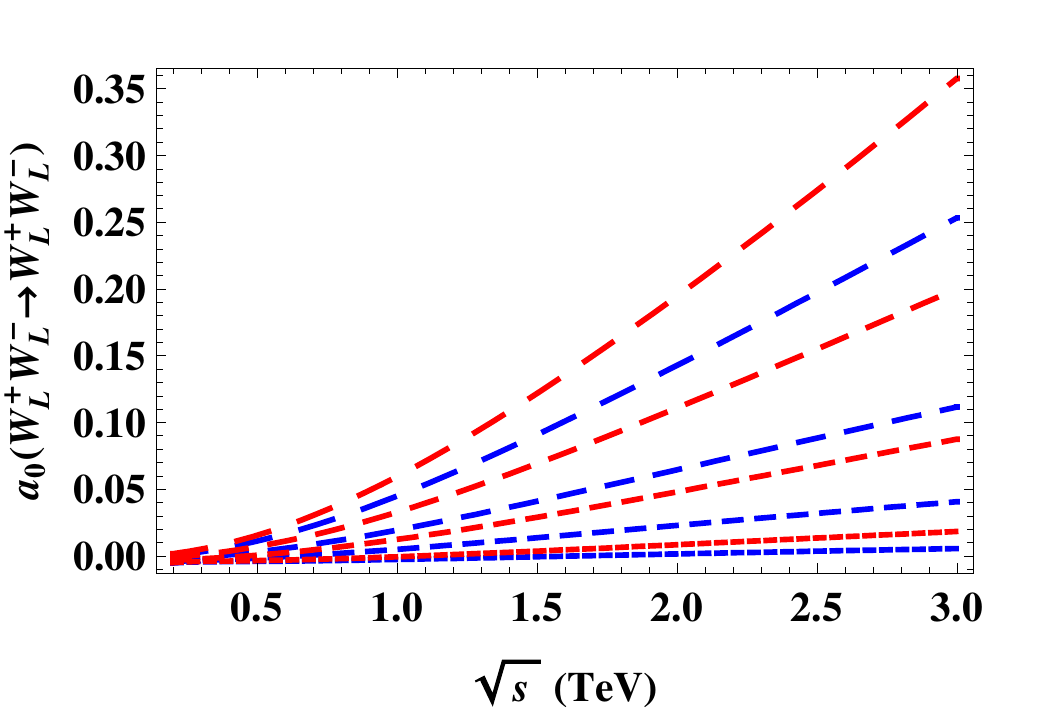}
\includegraphics[width=2.3in]{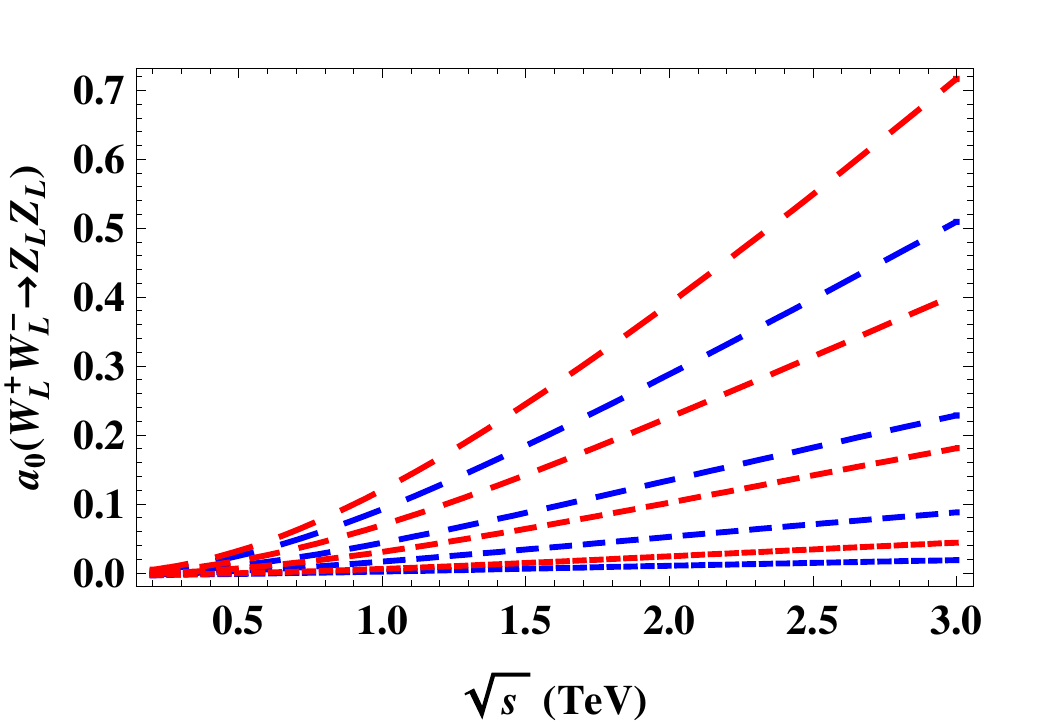}
\includegraphics[width=2.3in]{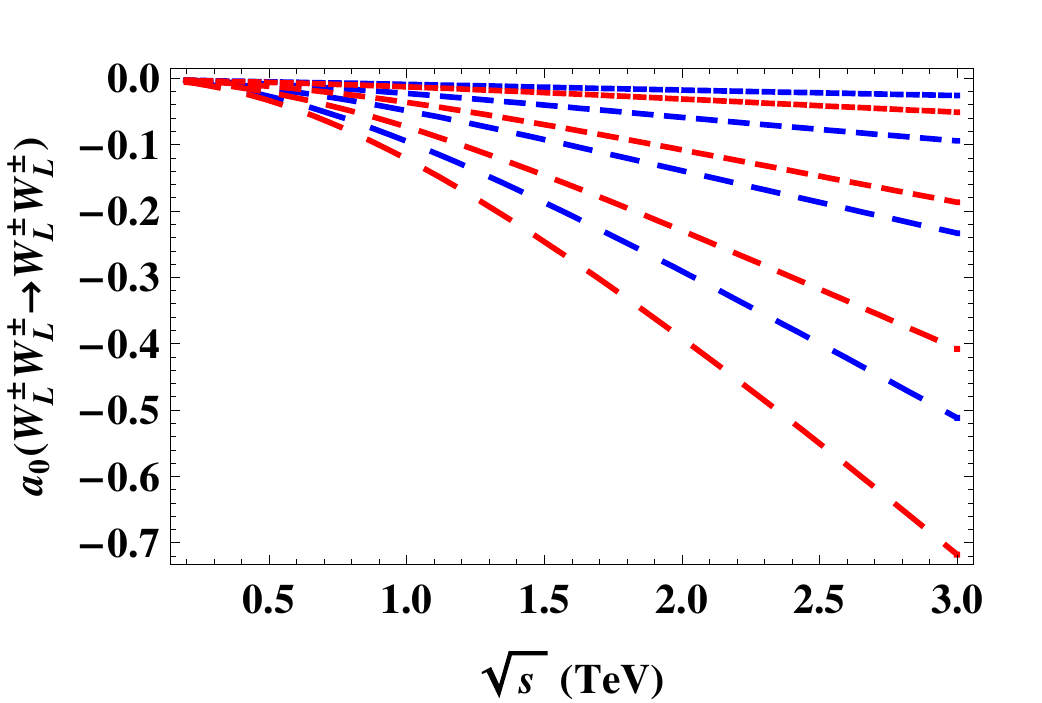}
\caption{$J=0$ partial wave projections of $W_L^+ W_L^-\to W_L^+ W_L^-$ (left), $W_L^+ W_L^-\to Z_L Z_L$ (center), and $W_L^\pm W_L^\pm\to W_L^\pm W_L^\pm$ (right) scattering amplitudes, as a function of the CM energy, for ${\cal S}=0.3$, $\beta=0$, and up to small electroweak corrections. Increasing dashing size corresponds to increasing values of $\tilde{g}$, with $\tilde{g}=2,4,6,8$. The blue curves are for $M_A=1.5$ TeV, whereas the red curves are for $M_A=2.0$ TeV. The amplitudes increase with $\tilde{g}$, in accordance with the behavior of $1-r_h^2-r_V^2$ in Fig.~\ref{fig:sumrule} (left).}
\label{fig:unitlong}
\end{figure}

Since larger values of $\tilde{g}$ give a larger coefficient of the term growing like $E^2$, we also expect the longitudinal weak boson scattering amplitudes to grow with $\tilde{g}$. This is shown in Fig.~\ref{fig:unitlong}, where the $J=0$ projections of $W_L^+ W_L^-\to W_L^+ W_L^-$ (left), $W_L^+ W_L^-\to Z_L Z_L$ (center), and $W_L^\pm W_L^\pm\to W_L^\pm W_L^\pm$ (right) scattering amplitudes are plotted as a function of the CM energy, up to small electroweak corrections, for ${\cal S}=0.3$ and $\beta=0$. Increasing dashing size corresponds to increasing values of $\tilde{g}$, with $\tilde{g}=2,4,6,8$. The blue curves are for $M_A=1.5$ TeV, whereas the red curves are for $M_A=2.0$ TeV. The amplitudes increase with $\tilde{g}$ and $M_A$, in accordance with the behavior of $1-r_h^2-r_V^2$ in Fig.~\ref{fig:sumrule}. We therefore expect signals, in the LHC processes $pp\to VVjj$, to grow accordingly.
\section{Technicolor signals in weak boson scattering}\label{sec:vector}
Since the work of~\cite{Bagger:1993zf,Bagger:1995mk}, many detailed studies have been dedicated to weak boson scattering in a strongly coupled regime
\cite{Hikasa:1991tw,Chanowitz:1993zh,Dobado:1999xb,Oller:1999me,Butterworth:2002tt,Chanowitz:2004gk,Englert:2008tn,Ballestrero:2008gf,Ballestrero:2009vw,
Ballestrero:2010vp,Ballestrero:2011pe,Ballestrero:2012xa}. In the absence of a light Higgs, the LHC, running at 14 TeV, would very probably be able to determine whether the weak bosons interact strongly at high energies~\cite{Ballestrero:2008gf,Ballestrero:2009vw,Ballestrero:2010vp,Ballestrero:2011pe}. See also~\cite{Ballestrero:2012xa} for the analysis at 7, 8, and 10 TeV. In the presence of a light TC Higgs the prospects for discovering strong dynamics trough weak boson scattering are substantially unchanged -- as unitarization still occurs in a strong regime -- but may require larger amounts of luminosity. Weak boson scattering, in a strongly coupled regime and a generic composite Higgs, has been investigated in~\cite{Cheung:2008zh,Ballestrero:2009vw,Ballestrero:2010vp}.

As motivated in Sec.~\ref{sec:intro}, a light composite Higgs is expected to show up in models of near-conformal TC, like the ``vanilla TC'' model introduced in Sec.~\ref{sec:unit}. This also features a near degenerate set of spin-one vector and axial isotriplets. The Higgs and the vector technimesons unitarize the longitudinal weak boson scattering amplitudes, leaving a footprint of strong dynamics in $p p\to V V j j$ processes. In this section we inquire on the possibility of uncovering such a footprint in the high energy region. For this purpose we consider the most relevant totally leptonic channels, in which both weak bosons decay into leptons: $jjW^+W^-\ra jj\ell^+\ell'^-\nu\bar{\nu}$, $jj W^\pm W^\pm\to jj \ell^\pm\nu \ell^\pm\nu$, $jjZZ\ra jj\ell^+\ell^-\nu\bar{\nu}$, and $jj W^\pm Z\to jj 3l\nu$. These channels give a cleaner signature than semi-leptonic channels\footnote{We did not consider $ZZ$ to four leptons due to its low cross section.}. Two perturbative orders contribute to these channels at the LHC: the purely electroweak $\ordEW$, which includes signal and electroweak background, and $\ordQCD$, which corresponds to the QCD background. Moreover, in the $jjW^+W^-\ra jj\ell^+\ell'^-\nu\bar{\nu}$ channel $t\bar{t}$+2 light jets production gives relevant contribution to the background. This occurs when the $b$-quarks are lost in the beam pipe or have too low energy, and has been taken into account.

Our study is based on the complete simulation of these processes at the parton level, using Monte Carlo techniques. Samples for the $\ordEW$ TC signal have been generated using the {\sc MadGraph/MadEvent}~\cite{MadeventPaper} framework. The TC model has been implemented in FeynRules~\cite{Christensen:2008py} and used for phenomenological studies in~\cite{Belyaev:2008yj,Andersen:2011nk,Frandsen:2011hj}. The processes have been generated using the decay chain technique, in which  the final state fermions are decay products of the weak bosons. The $\ordQCD$ QCD background and the $\ordEW$ SM samples have been produced using the {\sc phantom} program~\cite{Ballestrero:2007xq}, with the complete $2\ra 6$ set of diagrams. For $t\bar{t}+2j$ we have used {\sc MadGraph}, relying also on the decay of the top quarks.

~\tbn{tab:cutsbasic} features a set of basic cuts to be employed for this kind of analysis. It includes basic detector coverage (e.g.: $|\eta(\ell^\pm)| < 3.0$), minimum transverse momenta for all observed partons, and a cut against low energy jet activity $p_T (j) > 30 {\rm GeV}$, in order to avoid contamination from the underlying event. We require a minimum di-lepton invariant mass to avoid photon singularity, and, in addition, we consider two different ranges for the mass of the lepton pair in the channel with two neutrinos, $2j\ell^+\ell^-\nu\bar{\nu}$. We first select same-flavor charged leptons with a mass in the interval $76$ GeV $< M(\ell^+\ell^-) < 106$ GeV, corresponding to a lepton pair produced by the decay of a $Z$ boson. With the additional requirement of a large missing transverse momentum, $p_{Tmiss} >120\ {\rm GeV}$, we cut out $Z+$ jets production, and isolate the $2jZZ \ra 2j\ell^+\ell^-\nu\bar{\nu}$ channel. When the mass of the lepton pair is outside the quoted range, or the two oppositely charged leptons belong to different families, we consider the event to be a candidate for the $2jWW$ channel. Since we are interested in high energy scattering, we require $M(\ell^+\ell^-) > 250 \mbox{ GeV}$ for this kind of events. Furthermore, we employ additional cuts especially tailored to eliminate certain types of background. These include suppression of top production: $|M(j\ell\nu_{rec}) - M_{top} | > 15\ {\rm GeV}$ in $3\ell\nu+2$ jets channel, and $M(\ell j)>180\ {\rm GeV}$  in $jjW^+W^-\ra jj\ell^+\nu\ell^-\nu$, in which the top momenta cannot be reconstructed.

\begin{table}[t!]
\begin{center}
\begin{tabular}{|l|}
\hline
{\bf Basic Cuts} \\
\hline
$p_T(\ell^\pm) > 20 \mbox{ GeV}$ \\
\hline
$|\eta(\ell^\pm)| < 3.0$ \\
\hline
$M(\ell^+\ell^-) > 20 \mbox{ GeV}$\\
$M(\ell^+\ell^-) > 250 \mbox{ GeV}$ \quad ($jjW^+W^-$)\\
$76$ GeV $< M(\ell^+\ell^-) < 106$ GeV \quad ($jjZZ$)\\
\hline
$p_T(j) > 30 \mbox{ GeV}$ \\
\hline
$|\eta(j)| < 6.5$ \\
\hline
$|M(j\ell\nu_{rec}) - M_{top}| > 15 \mbox{ GeV}$ ($3\ell\nu + 2j$)\\
\hline
$M(\ell j)>180\GeV$ ($jjW^+W^-$)\\
\hline
$p_{Tmiss} >120\GeV$ ($jjZZ$)\\
\hline
\end{tabular}
\caption{Basic cuts imposed to isolate highly energetic $VVjj$ events. See text for details.}
\label{tab:cutsbasic}
\end{center}
\end{table}

In~\tbn{tab:extracuts} we enlist a set of kinematical cuts, applied for each channel, whose purpose is to enhance the discrepancy between the TC scenarios and the SM predictions, and to improve the signal to background ratio. For the \emph{tag}-jets we require high energy, large separation, and localization in the forward/backward regions. The weak bosons and their decay products are required to be central, have a large $p_T$, and be back-to-back in the azimuthal plane. We also require little color activity in the central region. See~\cite{Ballestrero:2010vp, Ballestrero:2011pe} for a detailed analysis of cut selection in weak boson scattering.

\begin{table}[t!]
\begin{center}
\begin{tabular}{|l|c|c|c|c|}
\hline
			&$jjW^+W^-$& $jjZZ$ & $jj\ell^\pm\nu \ell^\pm\nu$
			                       & $3\ell\nu$\\
\hline
$|\eta(\ell^\pm)|<$ 	& 2	& 3	& 3    & 2   \\
\hline
$M(j_fj_b)>$		& 1000	& 800	& 100  & 1000  \\
\hline
$|\Delta \eta(j_fj_b)|>$& 4.8	& 4.5	& 4.5  & 4.8   \\
\hline
$p_T(\ell\nu) >$	& 	&	&      & 200   \\
\hline
$p_T(\ell^+\ell^-)>$	& 	& 120 	&      & 200   \\		
\hline
$p_T(\ell) > $		& 20	& 20	& 50   & 20    \\
\hline
$min{p_T(j)} <$		&	&	& 120  &       \\
\hline
$E(j)>$			& 180 	& 	&      &       \\
\hline
$max|\eta(j)|>$		& 2.5 	&	& 2.5  &       \\
\hline
$|\eta(j)|>$		& 1.3 	& 1.9	&      & 1.2   \\
\hline
$|\Delta\eta(Vj)|>$	& 	&	&      & 1.5   \\
\hline
$\Delta\eta(\ell j)>$	& 0.8 	& 1.3	&      &       \\
\hline
$\Delta R (\ell j)>$	& 1 	& 	& 1.5  &       \\
\hline
$|\vec{p}_T(\ell_1)-\vec{p}_T(\ell_2)|>$
			& 220 	& 	& 150  &       \\
\hline
$|\vec{p}_T(\ell^+\ell^-)-\vec{p}_T^{\,miss}|>$
			& 	& 290 	&      &       \\
\hline
$\cos (\delta\phi_{\ell\ell})<$
			& -0.6 	& 	& -0.6 &       \\
\hline
\end{tabular}
\caption{Extra selection cuts tailored to enhance the discrepancy between the TC scenario and the SM predictions. See text for explanations.}
\label{tab:extracuts}
\end{center}
\end{table}

In order to compare our results with the ones of~\cite{Andersen:2011nk}, we set ${\cal S}=0.3$ and $\beta=0$. The Higgs mass is now set to $M_h=125$ GeV, which does not lead to dramatic changes with respect to the 200 GeV value chosen in~\cite{Andersen:2011nk}. In the $(M_A,\tilde{g})$ plane we take the values $M_A=1.5,\ 2.0$ TeV, and $\tilde{g}=4,\ 6,\ 8$, which are somewhat complementary to the accessible region for direct DY production of spin-one technimesons~\cite{Andersen:2011nk}. For these values we show, in~\tbn{tab:resww}, the effective cross section for the selected processes, after application of the kinematical cuts given in~\tbn{tab:cutsbasic} and~\tbn{tab:extracuts}. Here we analyze the large mass region, $M(\ell^+\ell^-)>300\ {\rm GeV}$, $M_T(ZZ)>300\GeV$, $M(\ell^\pm\ell^\pm)>300\ {\rm GeV}$, and $M(WZ)>600\ {\rm GeV}$, for $2jW^+W^-$, $2jZZ$, $2j\ell^\pm\ell^\pm\nu\nu$, and $2j3\ell\nu$, respectively. $M_T(ZZ)$ is the transverse mass of the $ZZ$-system:
\begin{equation}
\label{eqn:mtzz}
M_T^2(ZZ)=\left[\sqrt{M_Z^2+p_T^2(\ell\ell)}+\sqrt{M_Z^2+p_{Tmiss}^2}\right]^2-|\vec{p_T}(\ell\ell)+\vec{p}_{Tmiss}|^2.
\end{equation}
The numbers in the first row stand for $(M_A,\tilde{g})$. Notice how the signal increases with $\tilde{g}$ and/or $M_A$, in agreement with the behavior shown in Fig.~\ref{fig:unitlong}. Notice also an overall larger yield in TC, compared to the SM.

\begin{table}[t!]
\centering
\begin{tabular}{|c|c|c|c|c|c|c|c|c|}
\hline
 	   & (1.5,4)	  & (2.0,4) 	 &  (1.5,6)   & (2.0,6)    & (1.5,8)   & (2.0,8)    & SM    & $t\bar{t}jj$  \\
\hline
$2jW^+W^-$ & .187 & .192 & .209  & .226  & .257   & .291 & .179   &.173 \\
\hline
$2jZZ$	   & .0528 & .0553 & .0592 & .0694 & .0821  & .101 & .0540  & 0. \\
\hline
$2j\ell^\pm\ell^\pm\nu\nu$
	   & .101 & .107 & .113  & .131  &  .151  & .185  & .114   & 0.  \\
\hline
$2j3\ell\nu$
           & .0228 & .0238 & .0320  & .0346 & .0485  & .0547 & .017   & 0. \\
\hline	
\end{tabular}
\caption{Total cross section in $fb$ for each channel with the full set of cuts in \tbn{tab:cutsbasic} and \tbn{tab:extracuts}. We analyze the large mass region: $M(\ell^+\ell^-)>300\ {\rm GeV}$, $M_T(ZZ)>300\ {\rm GeV}$, $M(\ell^\pm\ell^\pm)>300\ {\rm GeV}$, and $M(WZ)>600\ {\rm GeV}$ for $2jW^+W^-$, $2jZZ$, $2j\ell^\pm\ell^\pm\nu\nu$, and $2j3\ell\nu$ respectively.}
\label{tab:resww}
\end{table}

\begin{figure}[t!]
\centering
\includegraphics*[width=0.45\textwidth,height=6cm]{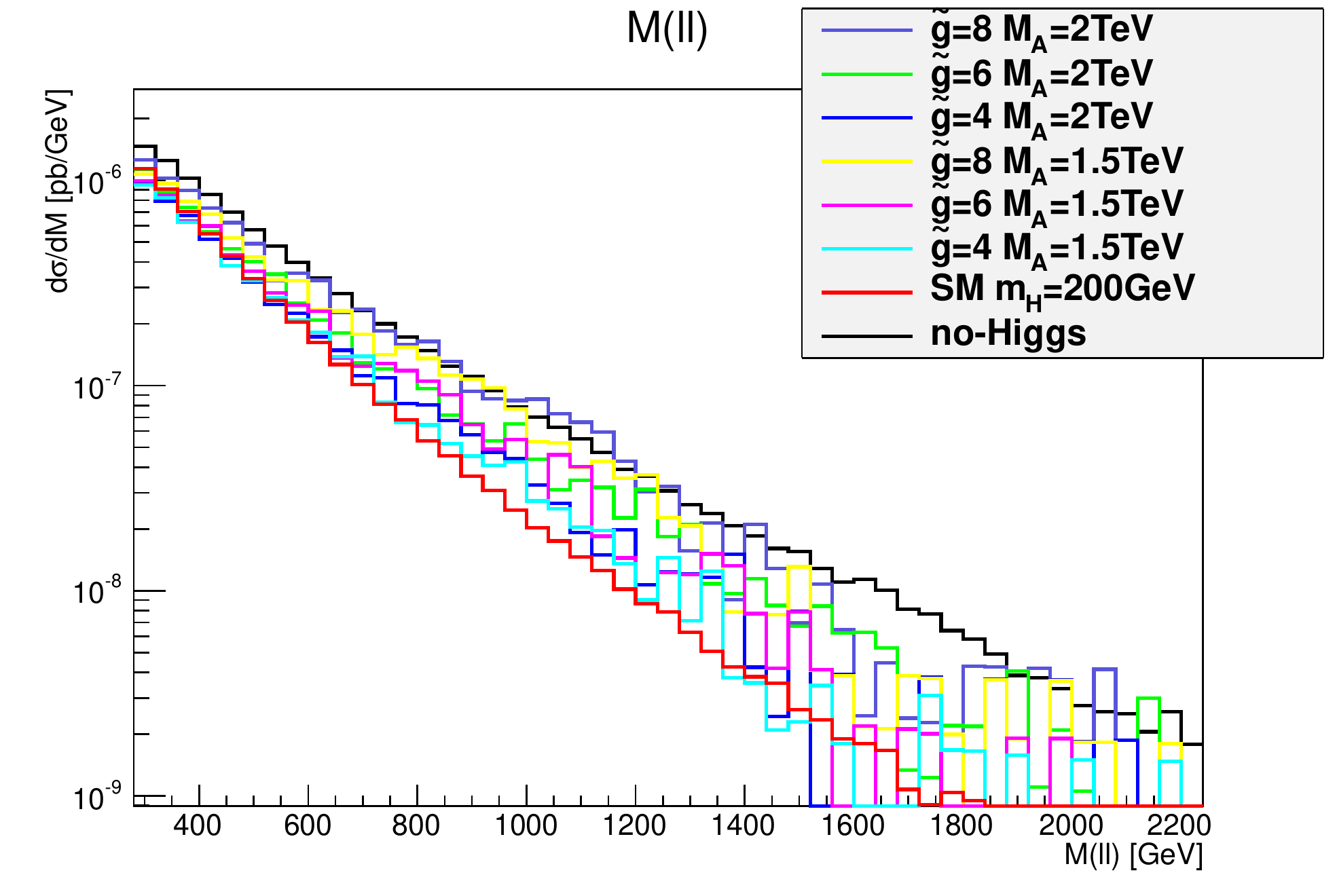}
\includegraphics*[width=0.45\textwidth,height=6cm]{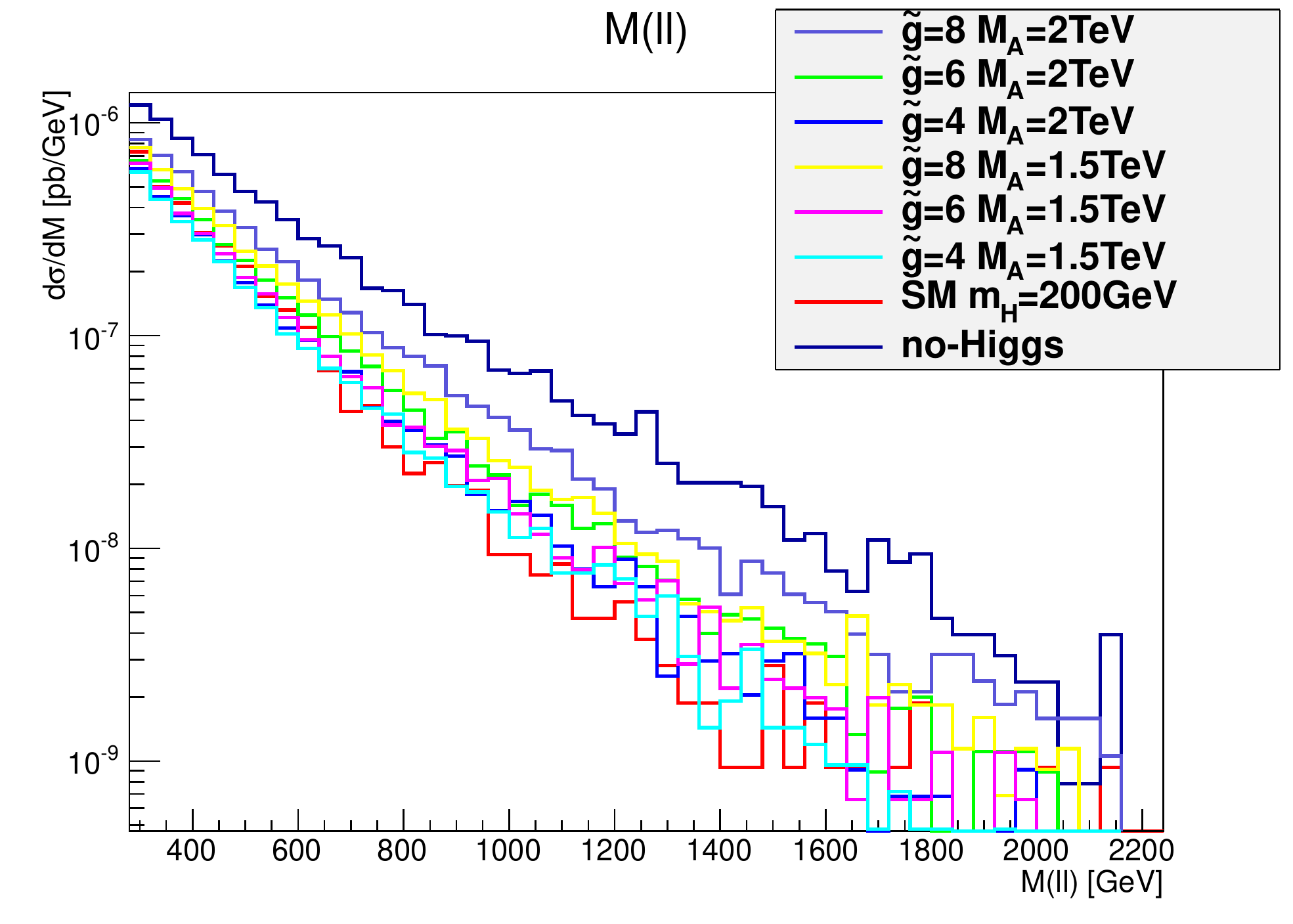}
\includegraphics*[width=0.45\textwidth,height=6cm]{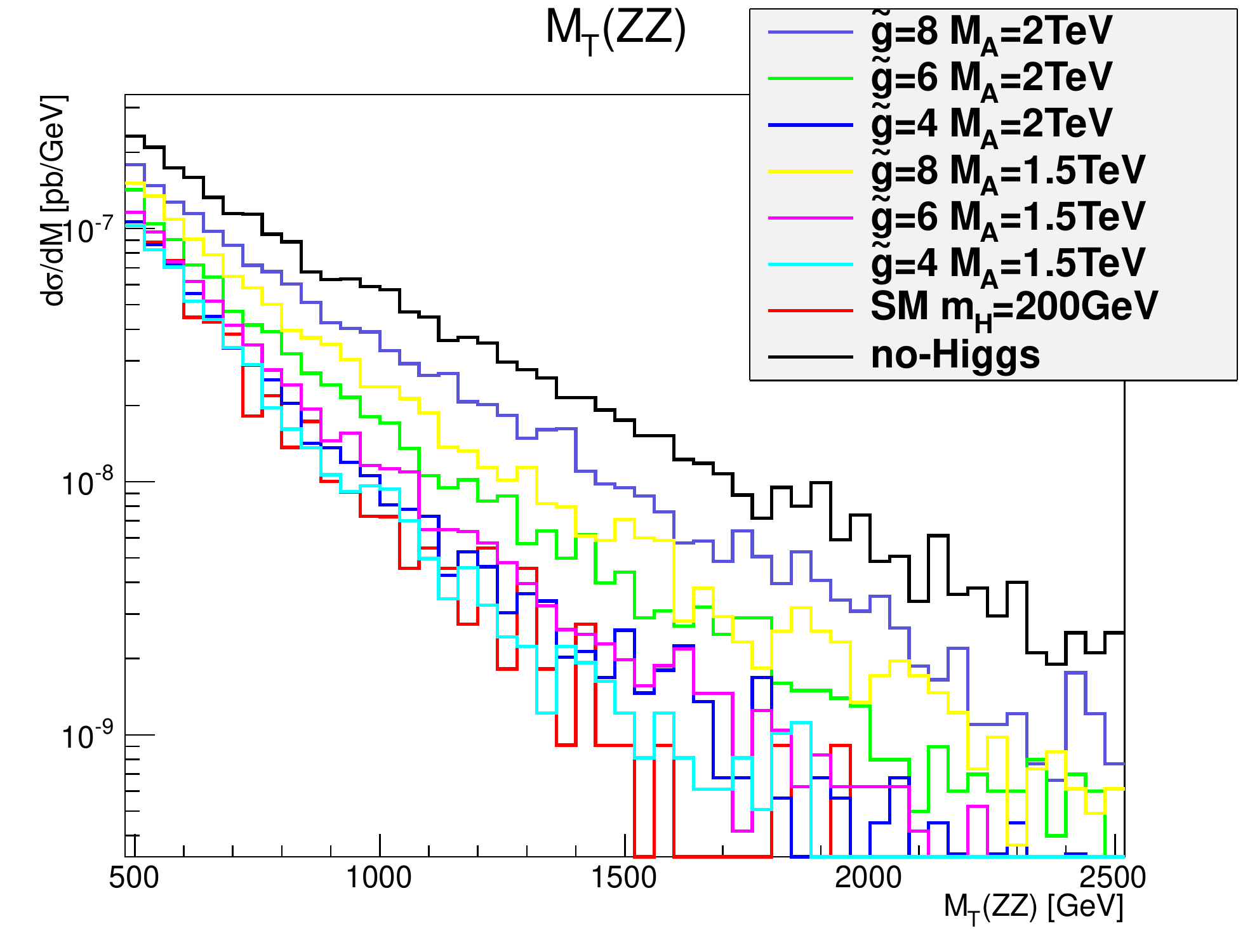}
\includegraphics*[width=0.45\textwidth,height=6cm]{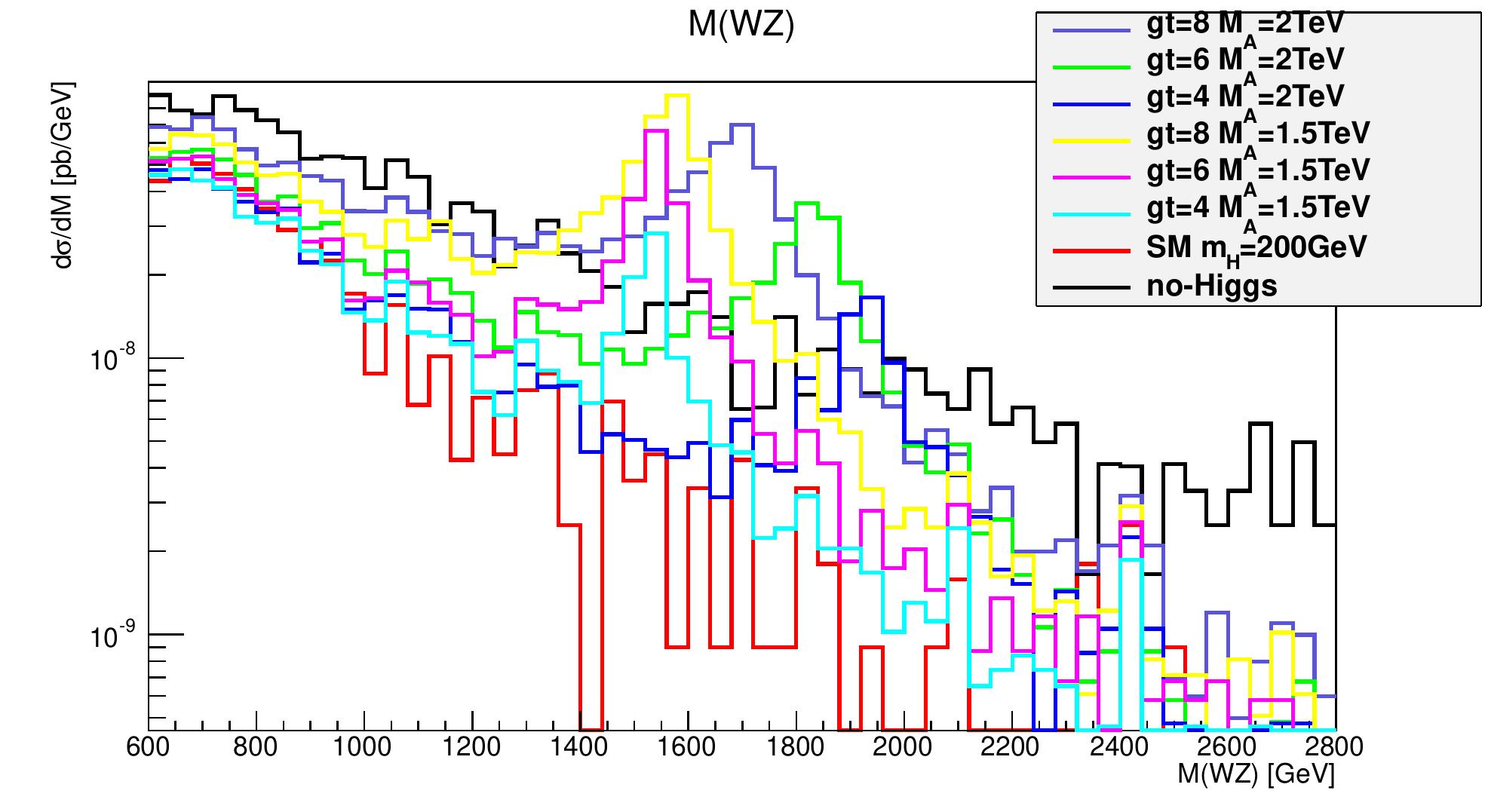}
\caption{Top-left:  di-lepton mass distributions for the $2jW^+W^-$ channel. Top-right: di-lepton mass distributions for the $2j\ell^\pm\ell^\pm\nu\nu$ channel. Bottom-left: transverse mass distributions of the $ZZ$-system for the $2jZZ$ channel. Bottom-right: invariant mass distributions of the $WZ$-system, with reconstructed neutrino, for the $2j3\ell\nu$ channel. For all of these distributions the cuts of \tbn{tab:cutsbasic} and \tbn{tab:extracuts} were imposed.}
\label{fig:VVjj}
\end{figure}

The channels $2jWW\ra 2j\ell\nu\ell'\nu$, either with opposite or same-sign leptons, present two neutrinos in the final state. As a consequence the reconstruction of the invariant mass of the complete di-boson system is not possible. Instead we must rely on the invariant mass of the two leptons in order to access the high energy region of the scattering process. The di-lepton mass distribution in the opposite-sign lepton channel is shown in~\fig{fig:VVjj} (top-left) for each of the TC scenarios considered, for the SM with a light Higgs, and also for the SM without the Higgs, for comparison. The corresponding distribution for the same-sign leptons channel is shown in~\fig{fig:VVjj} (top-right). In the $2jZZ\ra 2j\ell^+\ell^-\nu\nu$ channel, the presence of two neutrinos again prevents us from reconstructing the di-boson invariant mass. Nevertheless since the neutrinos are $Z$ decay products, it is possible to estimate the transverse mass of the $ZZ$-system, $M_T(ZZ)$, using the missing transverse energy, \eqn{eqn:mtzz}. The $M_T(ZZ)$ distribution for the $2j\ell^+\ell^-\nu\nu$ signal is shown in~\fig{fig:VVjj} (bottom-left). The only channel, among those considered, in which the di-boson mass can be more accurately reconstructed is $2jWZ\ra 2j3\ell\nu$. In this case the mass reconstruction can be achieved by using standard techniques to reconstruct the neutrino momentum~\cite{Ballestrero:2008gf}. The mass distribution of the $WZ$-system, with reconstructed neutrino momentum, is shown in~\fig{fig:VVjj} (bottom-right).

In all of the considered channels TC features an excess of events relative to the SM. The excess increases as $\tilde{g}$ and/or $M_A$ become larger, in agreement with the general behavior analyzed in Sec.~\ref{sec:unitscatt}, and summarized by the graphs of Fig.~\ref{fig:unitlong}. Phenomenologically this is a very interesting aspect, because it shows complementarity of our analysis with respect to the study of~\cite{Belyaev:2008yj,Andersen:2011nk,Frandsen:2011hj}. There, the discovery of TC relies on direct detection of DY-produced spin-one technimesons, in which the sensitivity drops with increasing $\tilde{g}$ and $M_A$. In the analysis of highly-energetic weak boson scattering the sensitivity actually increases with increasing $\tilde{g}$ and/or $M_A$. Of course it should be kept in mind that the mass of the spin-one technimesons cannot be too large, or else the effective theory undergoes an early loss of unitarity, and our analysis becomes inapplicable. However, as long as all weak-boson scattering amplitudes are unitary, our results hold. In fact, as already stressed in the introduction, in a strongly coupled regime the partial wave projections are always expected to saturate the unitarity bounds and oscillate between -1/2 and 1/2. We shall now quantify the potential for discovering near-conformal TC, through weak boson scattering, in the sum of all channels.

\subsection{Statistical combination and results}

In order to present our final results we have computed the probability to exclude the SM at the 95\% of confidence level, for each of the near-conformal TC scenarios we considered. To compute this probability we assume that the expected number of events observed in each scenario is given by the computed cross section times luminosity, $\sigma\times L$, and employ a Poisson distribution to account for statistical error. The contributions from $\ordEW$ and $\ordQCD$ cross sections are subject to parton distribution uncertainty and scale dependence, which have been taken into consideration by smearing the mean value of the poissonian by $\pm 30\%$.
QCD corrections to $VV$-scattering are around 10\%~\cite{Jager:2006zc,Jager:2006cp,Bozzi:2007ur,Jager:2009xx}, and PDF uncertainties around 5\% at the typical scale of these processes~\cite{Martin:2002aw}, hence the expected theoretical error are expected to be well within the value we assume.
On the other hand, $t\bar{t}+2$ jet backgrounds, which contribute to the $2jWW$ channel, are expected to be well measured from different regions of phase space, and extrapolated to the region of interest. Theoretical errors are therefore not expected to be an issue and we just assume a standard Poisson distribution in this case. We also assume that the theoretical errors are uncorrelated among different channels. We combine all the four considered channels, using as statistical test the \emph{likelihood ratio} distribution,
\begin{equation}
\label{eq:q}
Q(\vec{N};\vec{\langle N\rangle}_{TC},\vec{\langle N\rangle}_{SM})=\frac{
 \mathcal{P}(\vec{N};\vec{\langle N\rangle}_{TC})}{\mathcal{P}(\vec{N};\vec{\langle N\rangle}_{SM})}\,,
\end{equation}
where, within a specific model, $\mathcal{P}(\vec{N},\vec{\langle N\rangle})$ is the probability of obtaining the values $\vec{N}$ (number of events for the different channels) if the expectation values are $\vec{\langle N\rangle}$. As an example, in Fig.~\ref{fig:lnq} we show the probability distribution of $-2\ln Q$ for the combination of all four channels, comparing the $(M_A,\tilde{g})=(2\ {\rm TeV},8)$ near-conformal TC scenario with the SM. The probability to exclude the SM at the 95\% of confidence level can be extracted from the plot: the less the signal curve overlaps with the region on the right of the dashed vertical line, the larger the probability to exclude the SM. This probability is shown in~\tbn{tab:pbsm} for all TC scenarios we considered. As anticipated, the probability to exclude the SM is larger for larger values of $\tilde{g}$ and/or $M_A$. This demonstrates complementarity of this type of analysis with respect to the analysis based on DY searches of spin-one technimesons. Our results show that it is possible to exclude part of the parameter space already with $L=100\ifb$ at $14\ {\rm TeV}$. The analysis can be improved by adding additional decay channels for the weak bosons, and by exploring a larger portion of the parameter space.

\begin{figure}[t!]
\centering
\includegraphics*[width=0.52\textwidth,height=6cm]{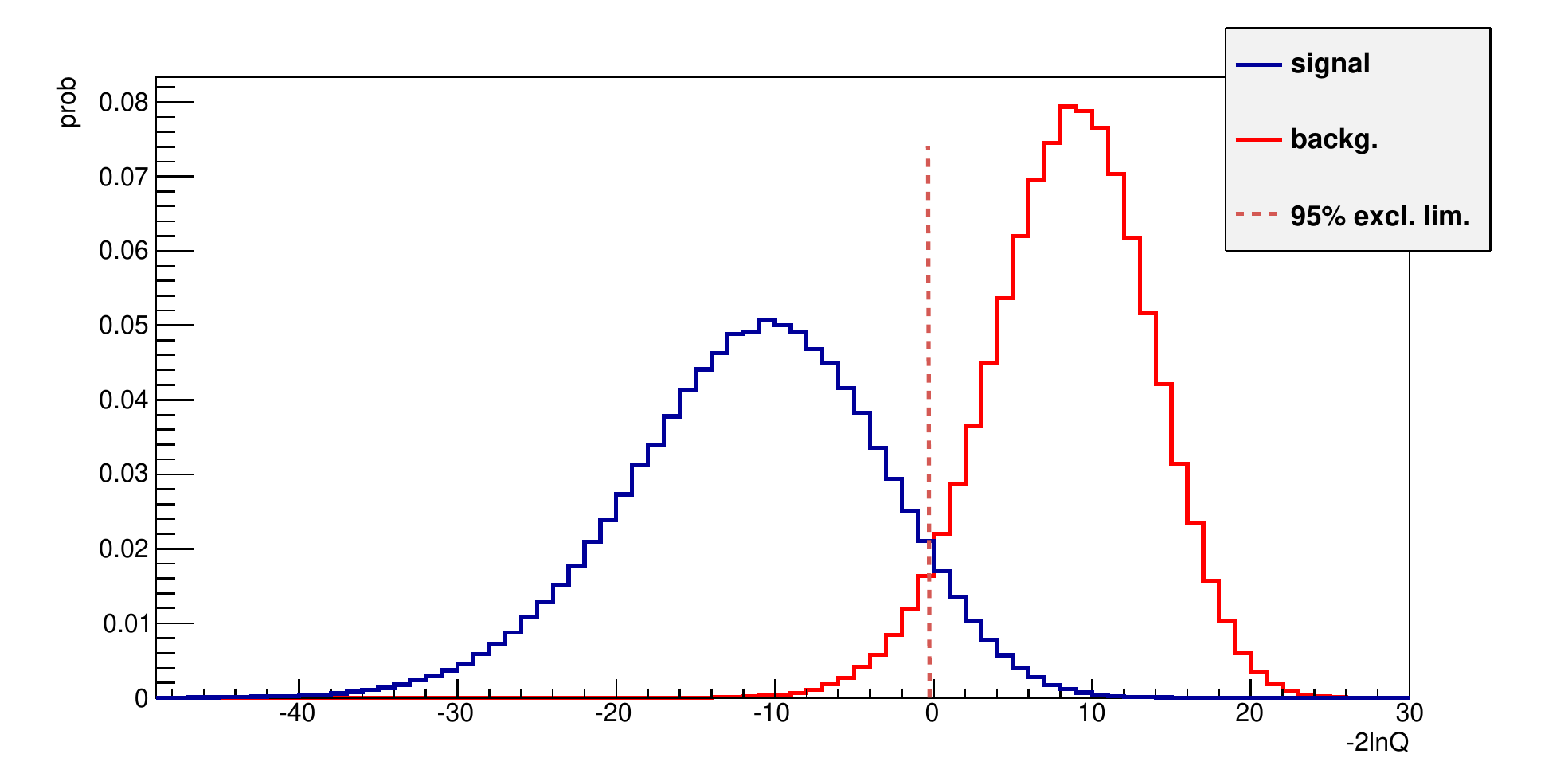}
\caption{One example of $-2\ln Q$ plot: combination of all channels for a comparison of the SM with the $(M_A,\tilde{g})=(2.0\ {\rm TeV},8)$ near-conformal TC scenario. The vertical dashed line indicates the value of $-2\ln Q$ for which the SM can be excluded at the 95\% C.L.: the less the signal curve overlaps with the region on the right of the dashed vertical line, the larger the probability to exclude the SM.}
\label{fig:lnq}
\end{figure}

\begin{table}[ht!]
\centering
\begin{tabular}{|c|c|c|c|c|c|c|}
\hline
	 & (1.5,4) &  (2.0,4) & (1.5,6) & (2.0,6) & (1.5,8) & (2.0,8)   \\
\hline
$50\ifb$ & 7.09 & 7.91 & 17.13 & 25.51 & 53.00 & 77.06 \\
$100\ifb$& 7.84 & 10.14  & 23.29 & 37.19 & 73.83 & 93.02 \\
$200\ifb$& 11.10 & 13.15 & 35.41 & 53.08 & 90.01 & 98.96 \\
$400\ifb$& 11.18 & 16.19 & 47.43 & 70.89 & 97.67 & 99.94 \\
\hline
\end{tabular}
\caption{Probability (in \%) to exclude the SM at the 95\% of confidence level, at 50, 100, 200 and 400 $\ifb$ of integrated luminosity, for different $(M_A,\tilde{g})$ near-conformal TC scenarios ($M_A$ values are shown in TeVs). These probabilities result from a statistical combination of the four considered channels, assuming uncorrelated theoretical errors.}
\label{tab:pbsm}
\end{table}
\section{Conclusions}\label{sec:conclusions}
It is possible that the recently observed boson at 125 GeV is a composite scalar singlet of a near-conformal TC theory: a TC Higgs. The presence of a nearby fixed point reduces the mass of the TC Higgs, relative to QCD-like estimates, and quadratically divergent radiative corrections from the top quark lead to a further large reduction. If the near-conformal TC scenario is correct, then we expect also a near-degenerate set of spin-one vector and axial technimeson triplets to be in the spectrum. In~\cite{Belyaev:2008yj,Andersen:2011nk} the DY production of such resonances, at the LHC, was analyzed. The results are summarized in Fig. 1 of~\cite{Andersen:2011nk}: the LHC reach, at 13 TeV and 100 $\ifb$, is drastically reduced as the technimeson coupling $\tilde{g}$ is increased. The reason for this is that in VMD (which is typically assumed to hold) the SM fermion coupling with the spin-one TC resonances is reduced by a factor of $g/\tilde{g}$, where $g$ is the weak coupling. DY production also drops with increasing mass of the spin-one resonances, as obviously expected from kinematical reasons.

In this note we investigated signatures of near-conformal TC by analyzing highly energetic weak boson scattering amplitudes. The reason for this is that in a strongly coupled regime, and at high energies, the longitudinal weak boson scattering amplitudes are expected to saturate the unitarity bounds of Eq.~(\ref{eq:unitbound}). In a perturbative regime, in contrast, scattering amplitudes are expected to drop to negligible values at high energies. In near-conformal TC, at a few TeVs, the longitudinal weak boson scattering amplitudes are unitarized by the light TC Higgs and by the spin-one vector triplet. Since amplitudes are unitary but large, we expect enhanced signals in $pp\to VVjj$ processes, which can therefore be used to test the TC hypothesis. In contrast to DY production, the enhancement of weak boson scattering drops neither with increasing $\tilde{g}$ nor with increasing mass of the spin-one resonances, as the partial wave projections of the scattering amplitudes are always expected to saturate the unitarity bound. We actually find the signal for the $2jW^+W^-$, $2jZZ$, $2j\ell^\pm\ell^\pm\nu\nu$, and $2j3\ell\nu$ channels to increase with $\tilde{g}$ and $M_A$ ($M_A$ is the mass of the spin-one axial triplet, which, being near degenerate with the mass of the spin-one vector triplet $M_V$, can be conveniently used to parametrize the overall mass of the lightest states in the TC spin-one sector). This shows that TC searches in $pp\to VVjj$ channels are complementary to DY searches, which are most effective at small values of $\tilde{g}$ and $M_A$.

Our results are summarized in~\tbn{tab:pbsm}, in which we show the probability of excluding the SM at the 95\% C.L., for different near-conformal TC scenarios (different values of $\tilde{g}$ and $M_A$), and different integrated luminosities. For instance, if the TC scenario with $M_A=2.0$ TeV and $\tilde{g}=8$ is realized in Nature, then the LHC will be able to exclude the SM, with a 77\% probability, at the 95\% C.L., already for 50 $\ifb$ of integrated luminosity. Our analysis includes leptonic decays only, and can therefore be improved by including additional decay modes for the weak bosons. The analysis can also be refined by investigating a larger set of points in the parameter space.
The reach of this analysis goes to regions of the parameter space inaccessible to the usual DY production of resonances and is therefore an important tool for probing this type of theory.
\section*{Acknowledgments}
We would like to thank M.T. Frandsen for reviewing the manuscript. The work of R.F. is supported by the Marie Curie IIF grant.

\end{document}